\documentclass[twoside,twocolumn,9pt]{article}
\usepackage{extsizes}
\usepackage[super,sort&compress,comma]{natbib} 
\usepackage[version=3]{mhchem}
\usepackage[left=1.5cm, right=1.5cm, top=1.785cm, bottom=2.0cm]{geometry}
\usepackage{balance}
\usepackage{mathptmx}
\usepackage{sectsty}
\usepackage{graphicx} 
\usepackage{lastpage}
\usepackage[format=plain,justification=justified,singlelinecheck=false,font={stretch=1.125,small,sf},labelfont=bf,labelsep=space]{caption}
\usepackage{float}
\usepackage{subfiles}
\usepackage{fancyhdr}
\usepackage{fnpos}
\usepackage{caption}
\usepackage[english]{babel}
\addto{\captionsenglish}{%
  
}
\usepackage{tabularx}
\usepackage{array}
\usepackage{droidsans}
\usepackage{charter}
\usepackage[T1]{fontenc}
\usepackage[usenames,dvipsnames]{xcolor}
\usepackage{setspace}
\usepackage[compact]{titlesec}
\usepackage{hyperref}

\usepackage{epstopdf}

\definecolor{cream}{RGB}{222,217,201}

\begin{document}

\pagestyle{fancy}
\thispagestyle{plain}

\fancypagestyle{plain}{
\renewcommand{\headrulewidth}{0pt}
}
\fancyhf{}
\lfoot{\thepage}

\makeFNbottom
\makeatletter
\renewcommand\LARGE{\@setfontsize\LARGE{15pt}{17}}
\renewcommand\Large{\@setfontsize\Large{12pt}{14}}
\renewcommand\large{\@setfontsize\large{10pt}{12}}
\renewcommand\footnotesize{\@setfontsize\footnotesize{7pt}{10}}
\makeatother

\renewcommand{\thefootnote}{\fnsymbol{footnote}}
\renewcommand\footnoterule{\vspace*{1pt}%
\color{cream}\hrule width 3.5in height 0.4pt \color{black}\vspace*{5pt}} 
\setcounter{secnumdepth}{5}

\makeatletter 
\renewcommand\@biblabel[1]{#1}            
\renewcommand\@makefntext[1]%
{\noindent\makebox[0pt][r]{\@thefnmark\,}#1}
\makeatother 
\renewcommand{\figurename}{\small{Fig.}~}
\sectionfont{\sffamily\Large}
\subsectionfont{\normalsize}
\subsubsectionfont{\bf}
\setstretch{1.125} 
\setlength\bibsep{1pt}
\setlength{\skip\footins}{0.8cm}
\setlength{\footnotesep}{0.25cm}
\setlength{\jot}{10pt}
\titlespacing*{\section}{0pt}{4pt}{4pt}
\titlespacing*{\subsection}{0pt}{15pt}{1pt}



\makeatletter 
\newlength{\figrulesep} 
\setlength{\figrulesep}{0.5\textfloatsep} 

\newcommand{\topfigrule}{\vspace*{-1pt}%
\noindent{\color{cream}\rule[-\figrulesep]{\columnwidth}{1.5pt}} }

\newcommand{\botfigrule}{\vspace*{-2pt}%
\noindent{\color{cream}\rule[\figrulesep]{\columnwidth}{1.5pt}} }

\newcommand{\dblfigrule}{\vspace*{-1pt}%
\noindent{\color{cream}\rule[-\figrulesep]{\textwidth}{1.5pt}} }

\makeatother



\twocolumn[
  \begin{@twocolumnfalse}

\vspace{1em}
\sffamily
\begin{tabular}{m{2.5cm} p{13.5cm} }
 \includegraphics{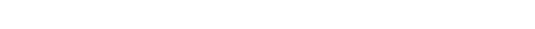} &\noindent\LARGE{\textbf{Phase behavior of active and passive dumbbells$^\dag$}} \\
\vspace{0.3cm} & \vspace{0.3cm} \\

 & \noindent\large{Nayana V\textit{$^{a}$}, Shiang-Tai Lin\textit{$^{b}$}, Prabal K Maiti\textit{$^{a}$}} \\
 \vspace{1cm}

\includegraphics{head_foot/LF.pdf} & \noindent\normalsize{Using molecular dynamics simulations, we report   phase separation in a 50:50 mixture of 'hot' and 'cold' dumbbells which interact by Lennard-Jones potential. The ratio of the temperature difference between hot and cold dumbbells to the temperature of cold dumbbells is a measure of the activity \(\chi\) of the system. From constant density simulations, we observe  that the 'hot' and 'cold' dumbbells  phase separate at  high activity ratio (\(\chi >5.80\)). The critical activity of dumbbells is higher compared to that of a mixture of hot and cold Lennard-Jones monomers \((\chi >3.44)\) . The extent of phase separation is higher for high density.  As activity increases, the cold dumbbells cohere to form a large cluster indicating increased phase separation which is quantified by an order parameter. The transfer of energy from hot dumbbells to cold dumbbells due to collisions and the change in the effective volume of hot dumbbells on phase separation affects the entropy of hot dumbbells. The  entropy of the hot dumbbells are  calculated using the two-phase thermodynamic (2PT) method and is compared with the entropy of dumbbells in an equilibrium system where all the dumbbells have same temperature as that of the 'hot' dumbbells in the non-equilibrium system. We find that the entropy of hot dumbbells is greater than the entropy of their equilibrium counterparts once the phase separation sets in.  The density of the phase separated system varies across the interface along the length of the simulation box where the cold dumbbells form dense clusters with density greater than the average density of the non-equilibrium system while the density of hot dumbbells is less than the average density of the non-equilibrium system. Also in the phase separated non-equilibrium system, the high kinetic pressure of hot dumbbells is balanced by the virial pressure of the cold dumbbells. We find that that phase separation pushes the cluster of cold dumbbells to have solid like ordering. Bond orientation order parameters reveal  that the  cold dumbbells form solid-like ordering consisting of predominantly FCC and HCP packing, but the individual dumbbells have random orientations.  The simulation of the non-equilibrium system at different ratios of number of hot dumbbells to cold dumbbells reveals that the critical activity decreases with increase in fraction of hot dumbbells.} \\
\end{tabular}
  \end{@twocolumnfalse} \vspace{0.8cm}
  ]
  

\renewcommand*\rmdefault{bch}\normalfont\upshape
\rmfamily
\section*{}
\vspace{-1cm}


\footnotetext{\textit{$^{a}$~Department of Physics, Indian Institute of Science, C. V. Raman Ave,Bengaluru 560012, India. E-mail: nayanav@iisc.ac.in, maiti@iisc.ac.in}}
\footnotetext{\textit{ {$^{b}$~Department of Chemical Engineering, National Taiwan University, No.1, Sec. 4 Roosevelt Rd. Taipei, Taiwan 10617 . E-mail:stlin@ntu.edu.tw}}}

\footnotetext{\dag~Supplementary Information available}


\footnotetext{}



\section{Introduction}
Active matter consists of individual units which are driven out of equilibrium as they consume energy from the surrounding medium or through internal mechanisms. Examples of active matter include swarms of fish, flocks of birds \cite{PhysRevLett.75.4326,PhysRevE.58.4828,PhysRevLett.100.218103,TONER2005170,PhysRevE.86.031918,PhysRevLett.75.1226,doi:10.1146/annurev-conmatphys-070909-104101,PhysRevLett.102.010602,LI2008699}, bacterial colonies, collection of cells in a tissue, components of cellular cytoskeleton \cite{Gupta2015,RevModPhys.85.1143,RevModPhys.69.1269,doi:10.1021/ar0001719,PhysRevLett.92.078101,HAGAN201674}etc.  Active matter can also be chemically synthesized and has far reaching applications in nano-scale active transport and drug delivery\cite{doi:10.1021/nn3028997}.  They exhibit fascinating properties like self-organization\cite{PhysRevE.99.032605,doi:10.1063/1.5079861,doi:10.1146/annurev-conmatphys-031214-014710,PhysRevLett.114.018301,PhysRevLett.110.055701,PhysRevLett.110.238301,PhysRevLett.123.228001,PhysRevLett.112.218304,PhysRevX.8.031080,C4SM00975D,Wensink14308,C6SM01760F,Paliwal_2018,C7SM01432E,Solon2015,Bertin_2009,PhysRevE.96.042605,C5SM02950C,PhysRevE.90.022131,PhysRevE.88.032102,doi:10.1063/5.0010043,C9SM00998A,D0SM00711K,article} and play very important role in biological systems.\\
In many studies on active particles, the activity of the constituents is represented by a vector model. In vector model the constituent particle(e.g. bacteria) absorbs energy to generate a force which propels it in a particular direction(usually along the body axis). The activity is a measure of how much energy  the constituent particle can take and convert into work (live bacteria have high activity but dead bacteria have zero activity). Recent studies have shown that for a system of  mixture of passive and active particles(run and tumble particles), in the limit when persistence length of active particles is comparable to the radius of the particle, the system of active and passive particle reduces to that of a system of 'hot' and 'cold' particles\cite{PhysRevFluids.2.043103}. Such representation of vectorial activity by different temperatures reduces a vector model into scalar model. Such scalar models also help in building analytically solvable microscopic models which gives  deep insight into physics of active matter\cite{PhysRevE.92.032118,PhysRevE.101.022120}.    In scalar model, the activity differences between two species in a system(e.g. live and dead) is represented by assigning two different temperatures or two different diffusivities\cite{PhysRevLett.116.058301} to two species. For example, \textit{Ganai et al}\cite{10.1093/nar/gkt1417} have proposed the use of scalar model to explain the radial distribution of chromatin inside nucleus. Phase separation in a system of equal mixture of  hot and cold Lennard-Jones(LJ) particles at different densities has been studied \cite{2019SMat...15.7275C}. Here the activity is defined as relative temperature difference between hot and cold particles. Recently, we have done simulation in a two temperature model of soft repulsive spherocylinders (SRS) and have shown emergence of various liquid crystalline order at packing fractions for which only isotropic phase is possible in equilibrium \cite{chattopadhyay2021heating}\\
In this paper, we study phase separation in system of hot and cold dumbbells\cite{doi:10.1063/1.1572811} which interact by Lennard Jones potential with aspect ratio(ratio of bond length to diameter) 1 . Dumbbells are not only a simplest way to model anisotropic particles but they also show interesting crystal structures. The dumbbells with low aspect ratio form plastic crystals with no orientational ordering\cite{doi:10.1063/1.468922} while the dumbbells with high aspect ratio form crystals with both translational and orientational order. In some recent works, colloidal dumbbells have been used to build periodic crystal structures which have partial photonic bandgap and are birefringent\cite{doi:10.1021/nn202227f}. Also, there are experimental works, which have  built chiral colloidal structures\cite{Zerrouki2008} with asymmetric dumbbells using magnetic fields. So the dumbbells have garnered much interest in nanotechnology of colloids.\\
In this paper, we study phase separation in  equal mixture of passive and active dumbbells. The active dumbbells are connected to a thermostat at high temperature whereas passive dumbbells are connected to a thermostat at low temperature to create difference in activity between the two components of the system. It is observed that such a simple system shows phase separation at sufficiently  high activity.\\
The paper is organized as follows: in the next section we give details of the simulation methodology. We next discuss the equilibrium properties of the dumbbells including the entropy calculation using both thermodynamic perturbation and 2-phase thermodynamic method. This is followed by the discussion of the results for non-equilibrium system of equal mixture of hot and cold dumbbells. The phase separation of hot and cold dumbbells is quantified by order parameter \(\phi\). We define another parameter \(\psi\) whose distribution P(\(\psi\)) helps in determining the critical activity.  The critical activity is determined by observing the emergence of bimodality in P(\(\psi\)). The transfer of energy from hot dumbbells to cold dumbbells due to collisions and the change in effective volume of hot dumbbells on phase separation affect the entropy of dumbbells. So we determine entropy difference \(\Delta\)S between hot dumbbells of non-equilibrium system  S\textsuperscript{neq} with entropy of corresponding equilibrium system  S\textsuperscript{eq}. We find that the entropy difference \(\Delta\)S jumps from negative to positive value with the onset of phase separation. Further, on performing cluster analysis of cold dumbbells, we find that with increasing activity \(\chi\) the number of cluster of cold dumbbells decrease, while the size of largest cluster increases. See section 3.2 for further details. The phase separation forces cold sub-system to have higher density and hot sub-system to have lower density than the average density of total system. The density, temperature and pressure variation along the length of the simulation volume is given in section 3.2.4 . In section 3.2.5, bond orientation order parameters are calculated which reveal that the cold cluster develop solid like order with predominantly fcc and hcp arrangement. Finally, we simulate the non-equilibrium system at different ratios of number of hot dumbbells to the number of cold dumbbells and find that critical activity decreases with increase in fraction of hot dumbbells. 
\section{Methodology and computational details}
\begin{figure}[t]
\centering
\includegraphics[width=1\columnwidth]{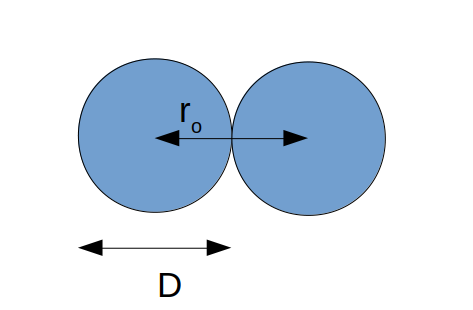} 
\caption{Schematic diagram of a dumbbell. D=\(\sigma\) is the diameter of each atom. r\textsubscript{o} is the distance between centers of two atoms in a dumbbell}
\end{figure}
We start with a system of 8000  atoms where two atoms are bonded  to form 4000 dumbbells. A schematic diagram of dumbbell is shown in Fig:1.  Interaction between the atoms of the dumbbells is described by Lennard-Jones potential
\begin{equation}
    V(r)=4\epsilon\bigg[\bigg(\frac{\sigma}{r}\bigg)^{12}-\bigg(\frac{\sigma}{r}\bigg)^6\bigg]
\end{equation}
 where $r$ denotes the distance of separation between any two
atoms, \(\sigma\) and \(\epsilon\) correspond to the diameter and strength
of the interaction potential of the atoms. We have used
the parameters of argon (\(\sigma\)= 3.405 Å, \(\epsilon\) = 0.238 kcal  mol\textsuperscript{-1} and
mass m = 39.948 g  mol\textsuperscript{-1}) to convert the LJ units to real units.
Two atoms within a dumbbell are held by a harmonic bond potential of the following form
\begin{equation}
    V(r)=k(r-r\textsubscript{0})^2
\end{equation}
where r\textsubscript{0} is the equilibrium bond length and $k$ is bond strength. In our simulations, we have taken r\textsubscript{0} to be  equal  to 
the diameter of the atom(\(\sigma\))  and $k$  equal to 3000\(\epsilon/\sigma^2\) as per the original LJ chain paper\cite{doi:10.1021/j100076a028}.  \\
The Nos\'e-Hoover thermostat with damping parameter 50\(\delta\)t is used to maintain the temperature of the dumbbells. The time step of integration was chosen 
to be \(\delta\)t = 1 fs for all the simulations reported in this work.  All the
simulations are performed using the LAMMPS\cite{PLIMPTON19951} molecular
dynamics package.\\
\section{Results and analysis}
\subsection{Equilibrium properties}
\begin{figure}[t]
\centering
\includegraphics[width=1\columnwidth]{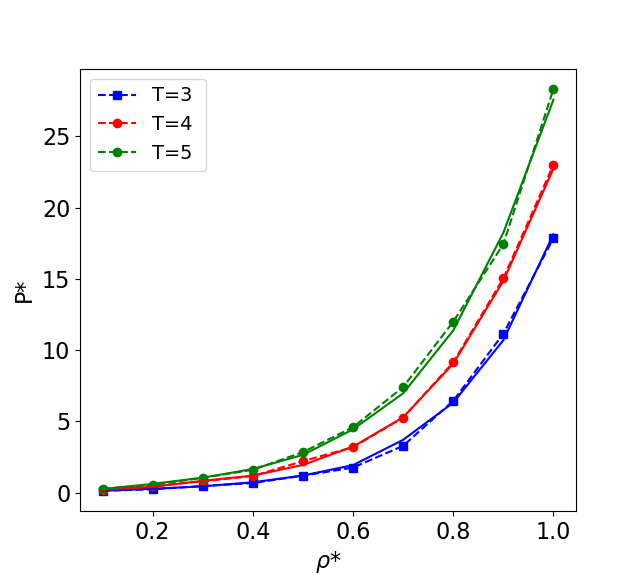} 
\caption{Pressure of dumbbells P* as a function of density \(\rho\)* at temperature T*=3,4 and 5. Dashed lines represent pressure obtained from perturbation theory. The solid lines represent the pressure obtained from MD simulations.}
\label{fig:fig1}
\end{figure}
In order to validate our system of dumbbells, we first calculate the equation of state and entropy of dumbbells. The temperature T*(k\textsubscript{b}T/\(\epsilon\)) and  density \(\rho*\)(\(\rho\)\(\sigma^3\)) are in reduced units. We run the MD simulation of dumbbells at T*=3,4,5(LJ units) at different densities to obtain the pressure of the system. The pressure obtained from simulation is compared with pressure obtained from Wertheim's Thermodynamic Perturbation theory\cite{doi:10.1021/j100076a028,doi:10.1063/1.2995990}. According to the Thermodynamic perturbation theory, the excess Helmholtz free energy of Lennard-Jones chain can be obtained from excess Helmholtz free energy of Lennard-Jones monomer fluid as below. 
\begin{equation}
  \frac{ A^n_e(\rho_n)}{N}= \frac{A_e(\rho)}{N} +T\frac{1-n}{n}lny(\sigma) 
\end{equation}
Here \(A^n_e\) is excess free energy of Lennard-Jones(LJ) chain composed of n monomers(for dumbbells n=2) , \(A_e\) is the excess free energy of Lennard-Jones monomer fluid, N is the total number of monomers and y(\(\sigma\)) is the pair correlation function of LJ monomer fluid at bond length \(\sigma\). Differentiating the above expression with respect to density we obtain the expression for pressure.
\begin{equation}
    P^n(\rho_n)= P(\rho)+\rho T\frac{1-n}{n}\bigg[1+\rho\frac{\delta lny(\sigma)}{\delta\rho}\bigg] 
  \end{equation}
  Here \(P^n\) is pressure of LJ chain fluid and \(P\) is the pressure of LJ monomer fluid. The pair correlation function y(\(\sigma\)) of LJ monomers is obtained as an empirical fitting function where it is expressed in powers of density and temperature. The empirical form of the pair correlation y(\(\sigma\)) is obtained from the work of Gubbin's et. al.\cite{doi:10.1021/j100076a028}. Pressure of the fluid of LJ monomer \(P\) is obtained from MD simulations. Pressure obtained using equation 4 (dashed lines) is plotted as a function of density in Fig:2. We can see that the pressure of dumbbells calculated from MD simulation (solid lines) matches closely with the values predicted by perturbation theory(dashed lines).\\
   \begin{figure}[t]
      
      \includegraphics[width=1\columnwidth]{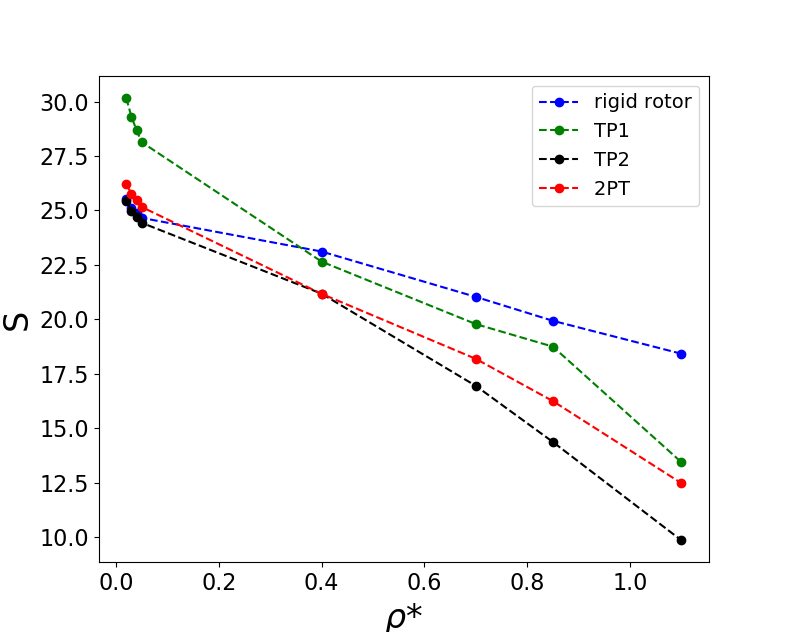}
      \caption{Entropy of dumbbell (S) as a function of density \(\rho*\) at temperature T*=1.8. Here 2PT stands for entropy by Two phase Thermodynamic Method , TP1 stands for entropy by Thermodynamic Perturbation with ideal monomer gas as reference and TP2 stands for entropy by Thermodynamic Perturbation with ideal diatomic gas(rigid rotor) as reference. .}
      \label{fig:my_label}
  \end{figure}
  
  Entropy is given by the negative derivative of free energy with respect to temperature. So differentiating equation 3 with respect to temperature gives the expression for entropy.   
  \begin{equation}
  \frac{ S^n_e(\rho_n)}{N}= \frac{S_e(\rho)}{N} +\frac{n-1}{n}\bigg[lny(\sigma)+T\frac{\delta lny(\sigma)}{\delta T}\bigg]
\end{equation}

where \(S^n_e\) is the excess entropy  of LJ chain, \(S_e\) is excess free energy of LJ monomer fluid. The excess entropy of LJ monomer fluid \(S_e\) is obtained from the modified  Benedict-Webb-Rubin equation of state\citep{doi:10.1080/00268979300100411}. To calculate the entropy of dumbbells from excess entropy of Lennard-Jones chain \(S^n_e\), we consider two references. The excess entropy of Lennard-Jones chain \(S^n_e\) is calculated with respect to ideal monoatomic gas (TP1) which is plotted in green dashed lines and  with respect to ideal diatomic gas (rigid rotor) (TP2) which is plotted in black dotted lines in Fig:3.\\
The entropy of dumbbells is also calculated from Two Phase Thermodynamic Method(2PT)\cite{doi:10.1063/1.1624057,doi:10.1021/jp103120q,C0CP01549K}. In 2PT method, first vibrational density of states of fluid is obtained from Fourier transform of velocity autocorrelation function. Then the vibrational density of states of the fluid is partitioned into a gas and solid like component. The gas like component is treated as a hard sphere gas while the solid component is treated as harmonic oscillator. This decomposition into solid and gas like fractions is performed from the value of vibrational density of states  of the liquid at zero frequency. Once the decomposition is done, by using the  weighting functions of harmonic oscillator and hard sphere gas, various thermodynamic properties like entropy and free energy can be calculated.
The entropy of dumbbell obtained from 2PT method as a function of density is plotted in Fig:3 red dashed line.\\
\begin{figure*}[!h]
\centering
\includegraphics[width=1\textwidth]{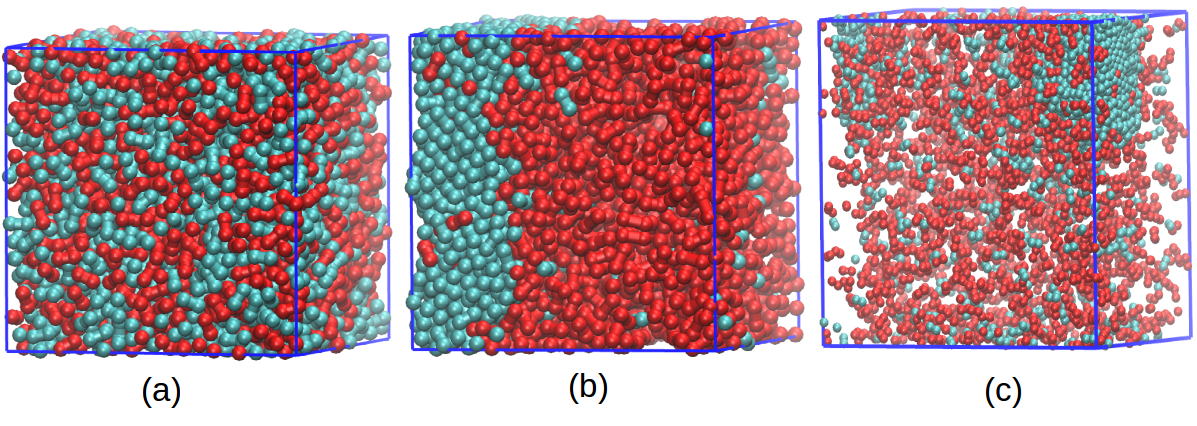} 
\caption{(a) Simulation at the end of 4ns when \(T\textsuperscript{h*}=T\textsuperscript{c*}=2\) and density is 0.8. We can see that hot(red) and cold(blue) particles are well mixed. (b) Simulation at the end of 1ns when \(T\textsuperscript{h*}=80\) and \( T\textsuperscript{c*}=2\) and density is 0.8. We can see that hot and cold dumbbells are clearly phase separated. (c) Simulation at the end of 1ns when \(T\textsuperscript{h*}=80\) and\( T\textsuperscript{c*}=2\) and density is 0.1. Here again, hot and cold dumbbells are phase separated. The snapshots were generated using VMD\cite{HUMP96} software.} \label{fig:fig1}
\end{figure*}
The dumbbell is approximated as a rigid rotor and the entropy of the rigid rotor as a function of density is also plotted in Fig:3. From Fig:3 we can see that entropy from 2PT falls between entropy calculated by Thermodynamic perturbation with ideal monoatomic gas(TP1) and idea diatomic gas(TP2) as reference.  Entropy from 2PT matches very closely with entropy of rigid rotor at very low density with a very small deviation. This small deviation is present because of the contribution of the vibration of dumbbells to the entropy. The entropy from 2PT follows the entropy of rigid rotor at low density  because the intermolecular interaction between dumbbells is negligible at low densities but deviates from the entropy of rigid rotor as density increases. So we believe that 2PT method can give the entropy of dumbbells effectively.

\subsection{Non-equilibrium Properties of Dumbbells}

In the system of 4000 dumbbells, to bring in 'activity' we introduce temperature difference between the dumbbells. Half of the dumbbells are labeled as 'cold'(passive) and are connected to a thermostat at low temperature and the remaining half dumbbells are labeled as 'hot'(active) and are connected to a thermostat at high temperature. Let T\textsuperscript{h}* and T\textsuperscript{c}* denote the temperatures of the ‘hot’ and
‘cold’ dumbbells respectively, in reduced units. Initially to equilibrate the system, both hot and cold dumbbells are maintained at same temperature at T\textsuperscript{h}* = T\textsuperscript{c}* = 2 and  at density \(\rho*\)=0.8. The simulation at constant temperature and volume  is run for 4 ns so that hot and cold dumbbells mix well (Fig:4a).\\

Initially, the temperature of the ‘hot’ dumbbells is increased 
from T\textsuperscript{h}* =2 to T\textsuperscript{h}*=5 and the simulation is run for 1 ns till the system reaches a steady state. Next, the temperature of hot dumbbells is increased from T\textsuperscript{h}* =5 to T\textsuperscript{h}* =10 so on up to T\textsuperscript{h}* =80  in steps of five(5, 10, 15...80), and the system is allowed to attain the steady state at each of the increased values
of T\textsuperscript{h}* and the simulation is run for 1 ns at each temperature. It is observed that even though the value of the thermostat temperature(T\textsuperscript{c*}) that is connected to the ‘cold’ dumbbells is left unchanged, there is an
increase in the temperature of the cold dumbbells due to heat
exchange from the ‘hot’ dumbbells through interactions. Hence,
they reach an ‘effective’ temperature \(T_{eff}^{c*} >T\textsuperscript{c*}\) . Here, the effective temperature is defined by the average kinetic energy of the dumbbells. In general,\\
         \(T\textsuperscript{h*}>T_{eff}^{h*}>T_{eff}^{c*}>T\textsuperscript{c*}\)\\
         
As the \(T\textsuperscript{h*}\) of the hot dumbbells is increased, cold dumbbells phase separate from the hot dumbbells. In Fig:4b, for density \(\rho*=0.8\) and T\textsuperscript{h}*=80, we observe that hot and cold dumbbells are phase separated.
         
To determine the extent of phase separation, we
divide the total volume of the simulation box into smaller sub-cells each having equal volume. To investigate phase separation at the sub-cell level we calculate the quantity \(\phi\) given by
\begin{equation}
\phi=\bigg|\frac{(n^h-n^c)}{(n^h+n^c)}\bigg|
\end{equation}
where n\textsuperscript{h} and n\textsuperscript{c} denote the number of hot and cold dumbbells respectively in each sub-cell . This quantity \(\phi\) is denoted as the order parameter and is then averaged over all the sub-cells and also over
all the steady-state configurations and is given by the following
equation\\
\begin{equation}
 \phi(T\textsuperscript{h}*)=\frac{1}{N_{cell}}\bigg<\sum_{i=1}^{N\textsubscript{cell}} \frac{|(n^h(i)-n^c(i))|}{(n^h(i)+n^c(i))}\bigg>\textsubscript{SS}
\end{equation}
where N\textsubscript{cell} is the total number of sub-cells and \(< >\textsubscript{SS}\) denotes that average is taken over all steady-state configurations. In our case N\textsubscript{cell}=125, the number is chosen such that there are enough number of dumbbells in each sub-cell.\\
Following the same procedure, simulations were also carried out at density \(\rho*\) of  0.5, 0.2, and 0.1.
\begin{figure}[h]
\centering
\includegraphics[width=1\columnwidth]{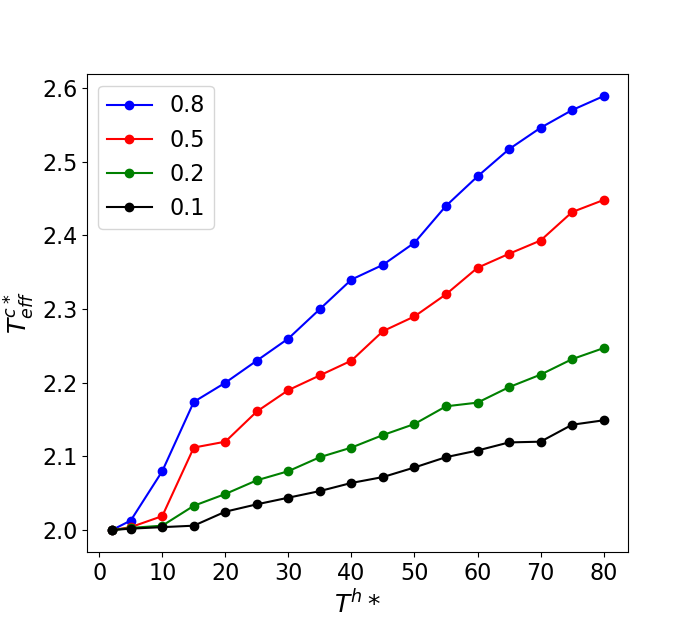} 
\caption{Plot of effective temperature of cold dumbbells \(T^{c*}_{eff}\) versus temperature of hot dumbbells \(T^{h*}\) at different densities. It is clearly seen that \(T^{c*}_{eff}\) increases with \(T^{h*}\) due to increased heat transfer from hot to cold dumbbells. Also, \(T^{c*}_{eff}\) increases with density because of the increase in frequency of collisions between hot and cold dumbbells. } \label{fig:fig1}
\end{figure}
\begin{figure}[h]
\centering
\includegraphics[width=1\columnwidth]{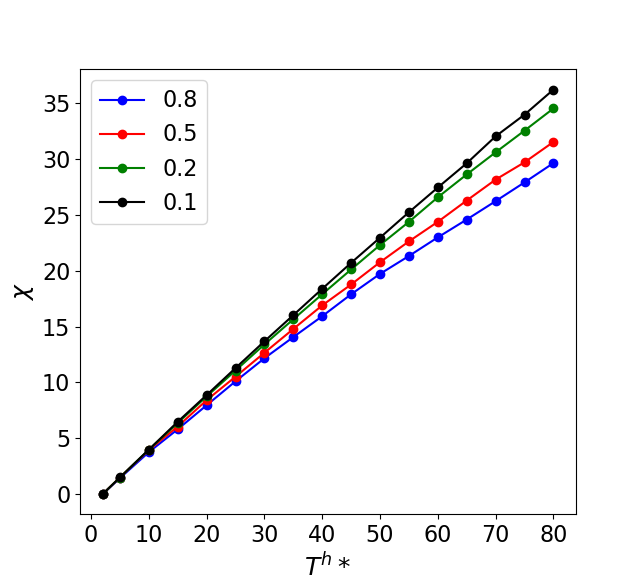} 
\caption{Plot of activity \(\chi\) versus temperature of hot dumbbells \(T^{h*}\) at different densities. In accordance with equation 8 activity \(\chi\) increases with increase in \(T^{h*}\). Also \(\chi\) increases with decrease in density because \(T^{c*}_{eff}\) increases as density increases(Fig:5).    } \label{fig:fig1}
\end{figure}
\begin{figure}[!h]
\centering
\includegraphics[width=1\columnwidth]{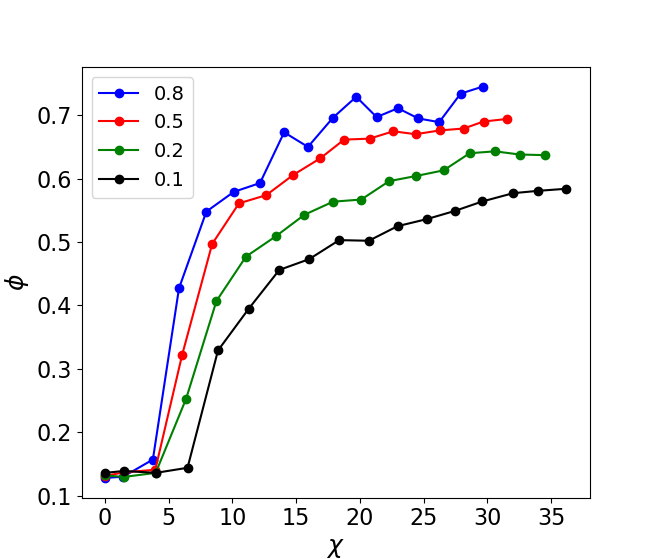} 
\caption{Plot of order parameter \(\phi\) versus activity \(\chi\) at different densities. Initially for low activity \(\chi\) order parameter \(\phi\) is small but as \(\chi\) increases  \(\phi\) rises and saturates to a high value indicating phase separation. Also value of \(\phi\) is greater for higher density indicating greater extent of phase separation at high densities.
} \label{fig:fig1}
\end{figure}
 \begin{figure*}[h]
      \includegraphics[width=1\textwidth]{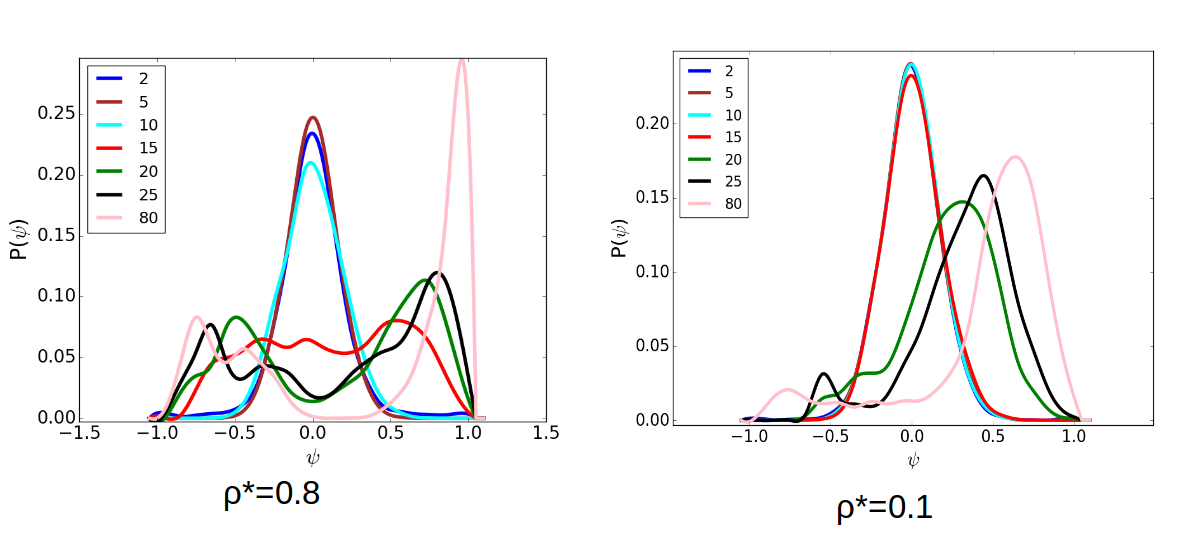}
      \caption{Distribution of \(\psi\) at density 0.8 and 0.1. The bimodality in distribution P(\(\psi\)) of  \(\psi\) occurs at T\textsuperscript{h*}=15 and 25 for density \(\rho*\) 0.8 and 0.1 respectively. These temperature of hot dumbbells T\textsuperscript{h*} correspond to the critical activity of 5.80 and 11.285. The distribution of \(\psi\) at density 0.5 and  and 0.2 is given in Fig:S3 in supplementary information.   }
      \label{fig:my_label}
  \end{figure*}

\subsubsection{Order parameter}
We define another parameter called activity
\begin{equation}
    \chi=\frac{(T_{eff}^{h*}-T_{eff}^{c*})}{T_{eff}^{c*}}
\end{equation} which is a measure of the difference in effective temperature of hot and cold dumbbells. It is the activity parameter \(\chi\) which is the measure of activity difference between hot and cold dumbbells.  When both hot and cold dumbbells are at same temperature, then activity \(\chi\) is 0. As the temperature of hot dumbbells is increased, the value of \(\chi\) also increases.\\ The plot of effective temperature of cold dumbbells \(T_{eff}^{c*}\) versus temperature of hot dumbbells T\textsuperscript{h*} is given in Fig:5 and plot of  activity \(\chi\) versus temperature of hot particles T\textsuperscript{h*} is given in Fig:6.\\

In Fig:5,  can see that as T\textsuperscript{h*} increases \(T^{c*}_{eff}\) also increases because of increased energy transfer from hot to cold dumbbells. Also as density increases,  \(T^{c*}_{eff}\)  increases because with increased density frequency of collision between hot and cold dumbbells increases. So the effective temperature of cold dumbbells \(T^{c*}_{eff}\) increases with  temperature of the hot sub-system and the density of the system.
In Fig:6 we can see that activity \(\chi\) increases with temperature of hot dumbbells \(T^{h*}\) in accordance with equation 8. The activity \(\chi\) also decreases with density because \(T^{c*}_{eff}\) which is in denominator in equation 8 increases with the density of the system.  \\

  When the value of activity \(\chi\) is low , the cold dumbbells form small clusters. As the activity is increased these small clusters join to form bigger clusters and finally at sufficiently high activity the phase separation becomes macroscopic where hot and cold dumbbells are clearly separated as shown in Fig:4b and Fig:4c.  \\

In Fig:7, a plot of order parameter \(\phi\) versus activity \(\chi\) at various densities is given.
It can be seen that, for very small initial activity \(\chi\) the order parameter \(\phi\)  remains small and constant.  As \(\chi\) increases, the value of \(\phi\) also increases and finally saturates to a large value. We can also see that value of \(\phi\) increases with density. So  higher density indicates higher extent of  phase separation. Also as density decreases, the point where \(\phi\) starts increasing is pushed to high activity. So as density decreases,  the initiation of  phase separation requires high activity and the extent of phase separation is also reduced. \\
To find out the critical activity at which  macroscopic phase separation of hot and cold dumbbells take place we plot the probability distribution of the parameter \(\psi\). The parameter \(\psi\) in i\textsuperscript{th} sub-cell is defined as 

\begin{equation}
 \psi=\frac{(n^h(i)-n^c(i))}{(n^h(i)+n^c(i))}
\end{equation}
The \(\psi\) takes negative values in the sub-cells where cold dumbbells dominate and positive values where hot dumbbells dominate. The value of \(\psi\) remains near zero where there is no phase separation and number of hot and cold dumbbells in a subcell are nearly equal. So we plot the distribution of \(\psi\) at different values of T\textsuperscript{h*} for each  density. Fig:8 gives the distribution \(P(\psi)\) for density 0.8 and 0.1. Refer to Fig:S3 in Supplementary Information for distribution \(P(\psi)\) for density 0.5 and 0.2.  As we increase the activity, the distribution of \(\psi\) becomes bimodal in shape when phase separation becomes macroscopic. The activity at which the distribution of \(\psi\) becomes bimodal is taken as critical activity. The bimodality in distribution of  \(\psi\) occurs at T\textsuperscript{h*}=15,20,20 and 25 for density \(\rho*\) 0.8, 0.5, 0.2 and 0.1 respectively. These temperatures of hot dumbbells T\textsuperscript{h*} correspond to the critical activity of 5.80, 8.41, 8.74 and 11.28. For a system of Lennard-Jones Monomer\cite{2019SMat...15.7275C} at density \(\rho*\)= 0.8, the critical activity is 3.44 but at the same density for a system of dumbbells, the critical activity is 5.80. So, the dumbbells start phase separating at a higher activity than the system of LJ monomers.    
\begin{figure*}[!h]
\centering
\includegraphics[width=0.8\textwidth]{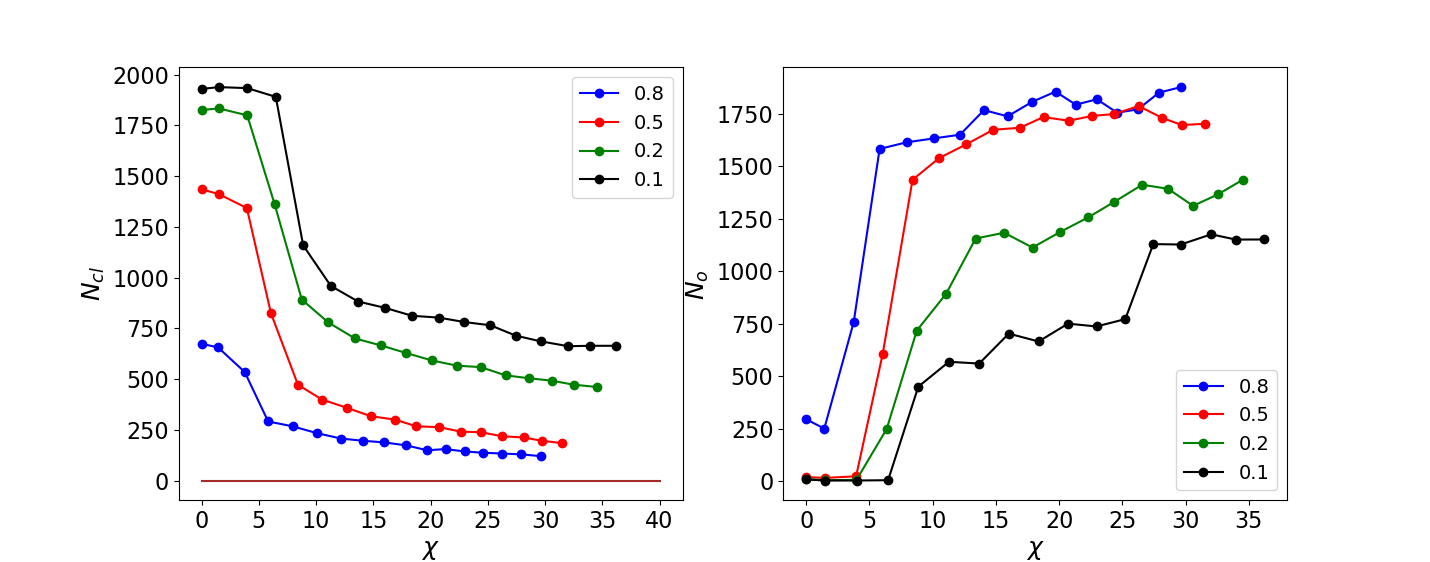} 
\caption{Plot of number of clusters of cold dumbbells N\textsubscript{cl} versus activity \(\chi\)(left) and plot of number of  cold dumbbells in the largest cluster N\textsubscript{0} versus activity \(\chi\)(right) . We can see that number of clusters of cold dumbbells N\textsubscript{cl} decreases with activity \(\chi\) as small clusters join to form large clusters. Complementary to N\textsubscript{cl} the number of cold dumbbells in the largest cluster N\textsubscript{0} increases with activity \(\chi\).} \label{fig:fig1}
\end{figure*}
\begin{figure}[!h]
 \includegraphics[width=1\columnwidth]{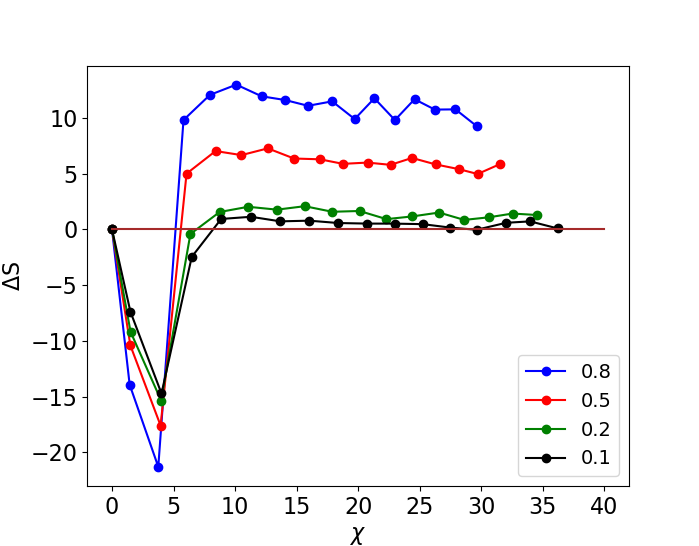}
 \caption{Plot of  entropy difference \(\Delta\)S= S\textsuperscript{neq}-S\textsuperscript{eq} versus activity \(\chi\) at different densities. We can see that \(\Delta\)S is negative before the system phase separates but jumps to a positive value once the phase separation sets in.} \label{fig:my_label}
  \end{figure}

\subsubsection{Cluster analysis}
As stated earlier, initially for low activity,  small clusters of cold dumbbells are formed. As activity is increased, these small clusters join to form larger clusters ultimately leading to phase separation. To study the number and size of clusters, we perform cluster analysis. In this work, we define a cluster using the following criterion: if the distance between two atoms of different dumbbells is within a specified cut off distance r\textsubscript{c} , then the dumbbells are considered to be part of the same cluster. The value of the cut off distance r\textsubscript{c} is determined from the position of the first peak in the radial distribution function calculated between the passive–passive dumbbells. Fig:14  shows the radial distribution function between cold dumbbells at \(\rho\)*=0.8 and T\textsuperscript{h*}=80. The value of r\textsubscript{c}=1.069\(\sigma\). Plot of  number of clusters of cold dumbbells N\textsubscript{cl} versus activity \(\chi\) is given in Fig:9(left). We can see that initially for small activity, when there is no phase separation number of clusters of cold dumbbells N\textsubscript{cl} is large and constant. As activity \(\chi\) is increased , phase separation sets in and the number of clusters of cold dumbbells decreases and saturates to a low value. Complementary to the number of clusters of cold dumbbells N\textsubscript{cl}, the size of the largest cluster increase with activity \(\chi\) as phase separation sets in. The number of cold dumbbells N\textsubscript{0} in the largest cluster is a measure of the size of the cluster.  Plot of number of cold dumbbells in the largest cluster N\textsubscript{0}  versus activity \(\chi\) is given in Fig:9(right).
Initially for low activity, when there is no phase separation, the size of largest cluster given by N\textsubscript{0} is small and constant. As the activity \(\chi\) increases, the the size of largest cluster grows and reaches a constant value. We can see that as density increases, number of dumbbells in the largest cluster N\textsubscript{0} and hence the size of the cluster also increases indicating higher extent phase separation. 

  \subsubsection{Entropy Calculation} 
  As mentioned earlier, there is a transfer of energy from hot dumbbells to cold dumbbells due to collisions between them. Also once the phase separation sets in, since the cold dumbbells form closely packed clusters as shown in section 3.2.2, the hot dumbbells have larger effective volume. These factors affect the entropy of hot dumbbells. Hence we compare the entropy of hot dumbbells with their equilibrium counterparts. Here we have calculated the entropy per dumbbell for the hot dumbbells in the non-equilibrium system of hot and cold dumbbells  which we call S\textsuperscript{neq}. It is compared with the entropy per dumbbell of an equilibrium system of dumbbells where the dumbbells have the same temperature as that of hot dumbbells T\textsuperscript{h*} of the non-equilibrium system and the density of equilibrium system is taken to be equal to the average density of non-equilibrium system, which is represented by S\textsuperscript{eq}. We use the 2-Phase Thermodynamic(2PT) \cite{doi:10.1063/1.1624057} method to calculate both entropy of hot dumbbells S\textsuperscript{neq} in non equilibrium system and entropy of dumbbells in corresponding equilibrium system S\textsuperscript{eq}.   
  We calculate the difference \(\Delta S\) defined as\\
  \begin{equation}
\Delta S=S\textsuperscript{neq} - S\textsuperscript{eq}
  \end{equation}
\\Plot of \(\Delta\)S versus \(\chi\) at various densities is given in Fig:10. 
We find that, initially for small activity \(\Delta\)S is negative, but as activity increases \(\Delta\)S jumps to a positive value.
There are two factors which determine value of \(\Delta\)S.\\
1. In non-equilibrium system, there is energy transfer from hot dumbbells to cold dumbbells. It decreases the entropy of hot dumbbells S\textsuperscript{neq} compared to that of equilibrium system S\textsuperscript{eq}.\\
  
2. When there is phase separation, cold dumbbells form dense clusters and hence effective volume for hot dumbbells increases. This increase in effective volume increases S\textsuperscript{neq} compared to S\textsuperscript{eq}. So we can see that initially for small activity where phase separation has not yet started, according to reason 1,  \(\Delta\)S is negative. As the activity is increased, phase separation takes place and reason 2 dominates reason 1 and \(\Delta\)S is positive. \\
In Fig:10, we can also observe that the value of  \(\Delta\)S increases with density.

\subsubsection{Density and temperature variation }
\begin{figure}[h]
\centering
\includegraphics[width=1\columnwidth]{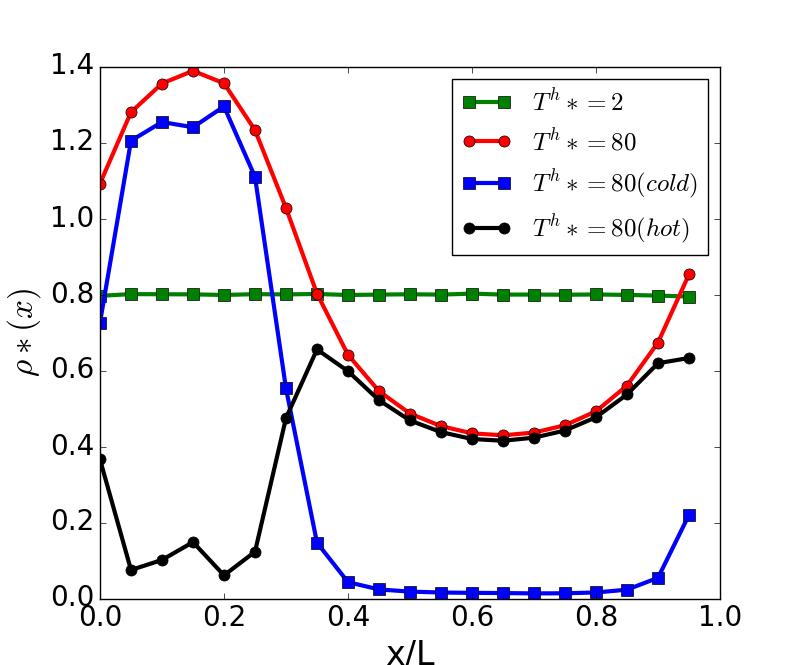} 
\caption{Plot of density variation \(\rho\)*(x) across the length of simulation box when total density \(\rho\)*=0.8 at T\textsuperscript{h*}=2 and 80 . We can see that density of cold dumbbells for T\textsuperscript{h*}=80  exceeds the average density 0.8 and the density of hot dumbbells decreases below average density. We can see that the interface is positioned near \(x/L\sim 0.3\) } \label{fig:fig1}
\end{figure}
\begin{figure}[h]
\centering
\includegraphics[width=1\columnwidth,]{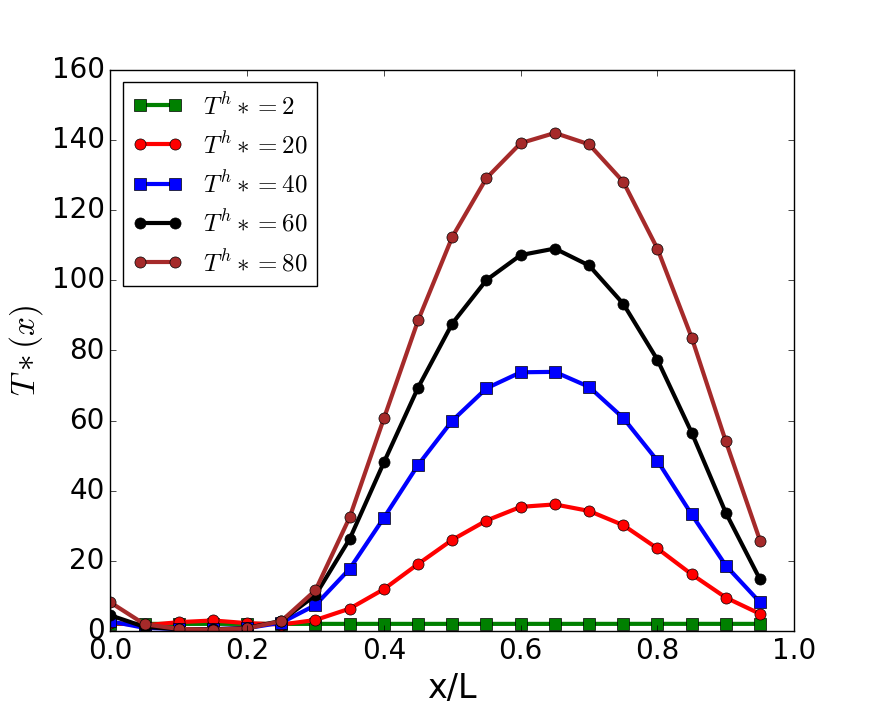}
\caption{Plot of temperature variation T*(x) across the length of simulation box for \(\rho\)*=0.8 at different T\textsuperscript{h*}. The temperature of the hot dumbbells (x/L>0.4) varies and decreases as we move towards the interface due to transfer of energy from hot to cold dumbbells.  } \label{fig:fig1}
\end{figure}
\begin{figure}[!h]
\centering
\includegraphics[width=1\columnwidth,]{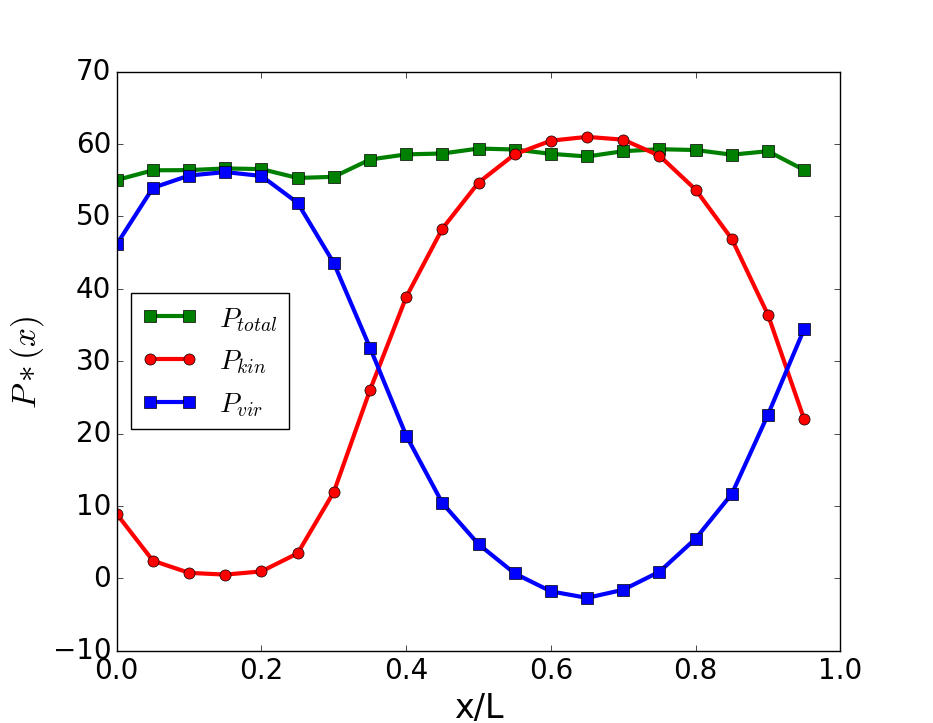}
\caption{Plot of pressure variation P*(x) across the length of simulation box at density \(\rho\)*=0.8 at  T\textsuperscript{h*}=80. The kinetic P\textsubscript{kin} and virial P\textsubscript{vir} contribution to the total pressure is plotted separately. The kinetic pressure of hot dumbbells is balanced by virial pressure of cold dumbbells. } \label{fig:fig1}
\end{figure}

In order to study how density and temperature varies along the length of the simulation box, the simulation box is divided in 20 sub-volumes along the direction phase separation takes place(say x axis). Then density  in i\textsuperscript{th} sub-volume is given by
\begin{equation}
    \rho(i)=\frac{n(i)}{V(i)}
\end{equation}
where n(i) is number of atoms of the dumbbells and V(i) is the volume of i\textsuperscript{th} sub-volume in reduced units.
Fig:11  gives plot of density \(\rho\)*(x) versus x/L for \(\rho\)=0.8 at T\textsuperscript{h*}=2 and T\textsuperscript{h*}=80, where L is the length of simulation box along x-axis. The density variation of hot and cold dumbbells has been plotted separately to  find the position of interface. From the plot, the interface seems to appear somewhere near x/L=0.3. We can see that density \(\rho*\)(x) of cold dumbbells before the interface(x/L=0.3) is much higher than that of average density 0.8, which indicates that cold dumbbells form densely packed clusters. Also the density of hot dumbbells (0.4<x/L<0.8)  is much lower than that of average density 0.8.\\
Each atom in a dumbbell has three  translational degrees of freedom. By equipartition theorem each degree of freedom contributes \(\frac{1}{2}k_bT\) of energy.
Hence temperature in i\textsuperscript{th} sub-volume is given by
\begin{equation}
    \frac{3}{2}k_bT(i)=\frac{1}{n(i)}\sum_{j=1}^{n(i)}\frac{1}{2}mv^2_j
\end{equation}

Fig:12 gives plot of temperature variation of dumbbells across the interface for different values of T\textsuperscript{h*}.
 We can see  from temperature variation that temperature of hot dumbbells is not uniform but decrease as they move towards the interface indicating energy transfer from hot to cold dumbbells.
Pressure profile along the interface is calculated from diagonal components of stress tensor.
\begin{equation}
    P(i)=\frac{1}{3*V(i)}[\sum_{j=1}^{n(i)}\frac{1}{2}mv^2_j+\sum_{j=1}^{n(i)}\vec{r_j}.\vec{f_j}]
\end{equation}
\begin{equation}
    P(i)=P_{kin}(i)+P_{vir}(i)
\end{equation}

Here \(\vec{r}_j\) is the position vector of j\textsuperscript{th} atom and \(\vec{f}_j\) is force on j\textsuperscript{th} atom due to all other atoms(includes force due to interaction with other dumbbells and the force due to bonding). The first term is the kinetic part of pressure P\textsubscript{kin} and second term is the virial part of pressure P\textsubscript{vir}. Fig:13 gives the pressure variation across the interface when \(\rho\)*=0.8 at  T\textsuperscript{h*}=80 . We have also plotted the kinetic and virial contribution to the pressure separately.  We can see that the pressure is almost constant along the length of simulation box. We can see that in the regions where the cold dumbbells dominate (x/L<0.3) the virial pressure P\textsubscript{vir} is very high and kinetic pressure P\textsubscript{kin} is low because of the low temperature of cold dumbbells. In the region where the hot dumbbells dominate (x/L>0.3) the kinetic pressure P\textsubscript{kin} is very high because of the high temperature of hot dumbbells. So, in the phase separated system the kinetic pressure of hot dumbbells is balanced by the virial pressure of cold dumbbells.

\subsubsection{Bond orientation order parameter}
\begin{figure}[!h]
\centering
\includegraphics[width=1\columnwidth]{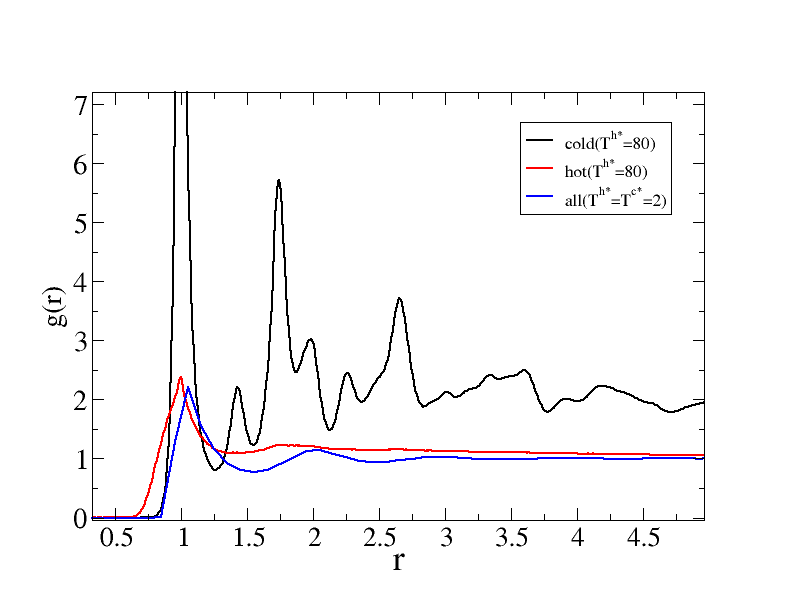} 
\caption{The g(r) of all dumbbells when T\textsuperscript{c*}=T\textsuperscript{h*}=2(blue) at \(\rho*\)=0.8 shows a liquid like structure with damping oscillations. The g(r) of the cold dumbbells(black) at \(\rho*\)=0.8 and T\textsuperscript{h*}=80 indicate solid like order but g(r) of hot dumbbells(red) at \(\rho*\)=0.8 and T\textsuperscript{h*}=80 indicate gaseous phase.} \label{fig:fig1}
\end{figure}
\begin{figure}[!h]
\centering
\includegraphics[width=1\columnwidth]{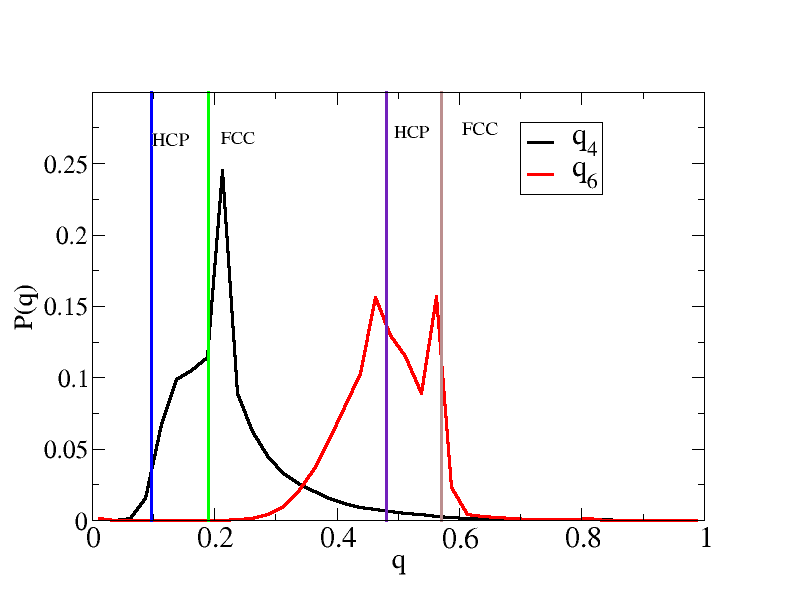} 
\caption{Distribution of bond orientation order parameters q4 and q6 of atoms of cold dumbbells when T\textsuperscript{h*}=80 at \(\rho*\)=0.8. From the distribution, we can see that the cold dumbbells have predominantly fcc and hcp arrangement} \label{fig:fig1}
\end{figure}
In Fig:14 we have radial distribution of all dumbbells(blue) at \(\rho*=0.8\) and T\textsuperscript{c*}=T\textsuperscript{c*}=2 which shows liquid like structure with damping oscillations. In the plot in Fig:14 we have radial distribution function g(r) between cold dumbbells(black) and radial distribution function g(r) between hot dumbbells(red)  for \(\rho*=0.8\) and T\textsuperscript{h*}=80. The g(r) of cold dumbbells show crystalline order but g(r) of hot dumbbells indicate a gaseous phase. 
From the density variation(shown in figure 11) , we can see that the cold dumbbells are forced to form dense clusters and from the radial distribution function (shown in figure 14) we can see that the clusters show crystalline order. So in the system of hot and cold dumbbells which is in liquid state when T\textsuperscript{c*}=T\textsuperscript{c*}=2, as the temperature of hot dumbbells is increased, the cold dumbbells have crystalline order while the hot dumbbells are in gaseous phase.\\

 In order to identify the local crystalline order we calculate Steinhardt bond order parameters. Two atoms of different dumbbells are considered to be neighbors if their distance is below 1.2\(\sigma\). The bond order parameters are defined as
\begin{equation}
    q_l(i)=\sqrt{\frac{4\pi}{2l+1}\sum_{m=-l}^{l}q_{lm}q^*_{lm}}
\end{equation}
where,
\begin{equation}
q_{lm}(i)=\frac{1}{N_b(i)}\sum_{j=1}^{N_b(i)}Y_{lm}(\theta_{ij}, \phi_{ij})
    \end{equation}
    N\textsubscript{b}(i) corresponds to the number of bonds for particle i, and Y\textsubscript{lm}
is the respective spherical harmonic function.
The ideal values of these parameters for various crystalline systems is given in table 1 .

\begin{table}
    \centering
    \begin{tabular}{|c|c|c|}
    \hline
       & q\textsubscript{4}    & q\textsubscript{6} \\ \hline
 fcc & 0.19 & 0.57 \\ 
 hcp & 0.097 & 0.48 \\ 
 \hline
    \end{tabular}
    \caption{Value of q\textsubscript{4} and q\textsubscript{6} for fcc and hcp crystals.}
    \label{tab:my_label}
\end{table}
\begin{figure*}[!h]
\centering
\includegraphics[width=1\textwidth]{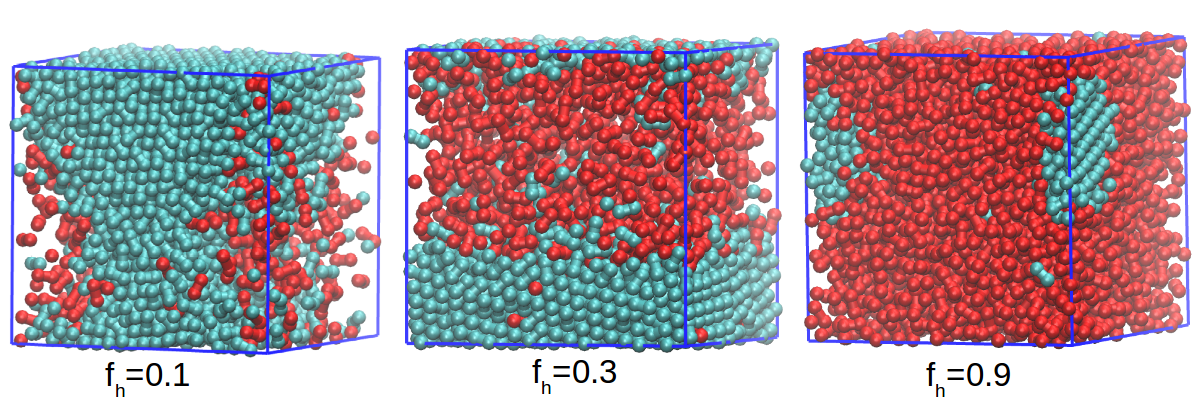} 
\caption{Snapshots of the non equilibrium system of hot and cold dumbbells at different fractions of hot dumbbells f\textsubscript{h} when \(\rho*\)=0.8 and T\textsuperscript{h*}=80 . We can see that hot and cold dumbbells phase separate at all given fractions of hot dumbbells f\textsubscript{h}. } \label{fig:fig1}
\end{figure*}

In Fig:15 distribution of bond order parameters q\textsubscript{4} and q\textsubscript{6} when T\textsuperscript{h*}=80 at \(\rho*\)=0.8 is given . We can see that the cold dumbbells form crystalline solid where the atoms of dumbbells are arranged so that they have predominantly face-centered cubic (fcc) and hexagonal-close packing (hcp) structure. Furthermore, each dumbbells point in different directions and does not posses orientational order. In an equilibrium system of dumbbells, at temperature T\textsuperscript{*}=2, above density \(\rho* = 1.08\) the dumbbells form disordered solid\cite{doi:10.1063/1.1465397} where atoms of dumbbells are in face centered cubic arrangement but the dumbbells have no orientational order and dumbbells point in different direction. In our non-equilibrium system of hot and cold dumbbells, we can see that the cold dumbbells form clusters of density much greater than \(\rho* = 1.08\) (Refer Fig:11) but the atoms of dumbbells show both hcp and fcc arrangements as obtained from bond order parameters. But the cold dumbbells of the solid cluster in the non-equilibrium system also do not posses orientational order like dumbbell crystal in the equilibrium system.  \\
\\
In the supplementary information (SI), we have compared velocity distribution of the whole hot subsystem with the velocity distribution of the hot dumbbells lying in middle of phase separated hot subsystem(refer figure S1 in supplementary information). We also have discussed the effect of damping parameter of thermostat on the non-equilibrium system of hot and cold dumbbells. Figure S2 gives plots of effective temperature of cold dumbbells \(T^{c*}_{eff}\) and activity \(\chi\) versus temperature of hot dumbbells \(T^{h*}\). We have also compared the pressure of different sub-cells with their equilibrium counter-parts which have same density and temperature.

\subsubsection{Effect of ratio of hot and cold dumbbells on phase separation}
All the results we have discussed so far correspond to 50:50 mixture of hot and cold dumbbells. We have also simulated the non-equilibrium system at different ratios of hot and cold dumbbells. The fraction of hot dumbbells f\textsubscript{h} in the system is given by
\begin{equation}
f_h=\frac{Number\, of\, hot\, dumbbells}{Total\, number\, of\, dumbbells}
\end{equation}
Following the same procedure mentioned in section 3.2, we have simulated  the mixture of hot and cold dumbbells for different fraction of hot dumbbells  f\textsubscript{h} 0.1, 0.3, 0.7 and 0.9 at density \(\rho\)* 0.8. The Fig:16 shows the snapshots of the system at T\textsuperscript{h*}=80 at f\textsubscript{h}  0.1, 0.3, and 0.9. We observe that for all given fraction of hot dumbbells f\textsubscript{h} the hot dumbbells and cold dumbbells phase separate. The effective temperature of cold dumbbells increases with increase in fraction of hot dumbbells in the system (See Fig:S4 and S5 in the supplementary information). We plot the probability  distribution P(\(\psi\)) of the parameter \(\psi\) defined in equation 9  for different fraction of hot dumbbells f\textsubscript{h} to identify the critical activity at which the phase separation of hot and cold dumbbells becomes macroscopic(See Fig:S6, S7, S8 and S9  in the SI for the plot of P(\(\psi\)) at different fraction of hot dumbbells f\textsubscript{h}). In Table:2 we give the value of critical activity for different fraction of hot dumbbells f\textsubscript{h}.
\begin{table}
    \centering
    \begin{tabular}{|c|c|}
    \hline
        f\textsubscript{h}    & critical activity \\ \hline
 0.1 & 15.28 \\ \hline
 0.3 & 8.14 \\ \hline
 0.5 & 5.80 \\ \hline
 0.7 & 3.56 \\ \hline
 0.9 & 3.30 \\ \hline

    \end{tabular}
    \caption{The critical activity of system of hot and cold dumbbells at different fraction of hot dumbbells f\textsubscript{h}. We can see that as the fraction of hot dumbbells f\textsubscript{h} increases the value of critical activity decreases}
    \label{tab:my_label}
\end{table}
From Table:2 we can see that as the fraction of hot dumbbells f\textsubscript{h} increases the value of critical activity decreases. So we can conclude that as the fraction of hot dumbbells is increased in the system, the hot and cold dumbbells start phase separating at smaller values of activity.

\section{Conclusion}
In a system of hot and cold dumbbells, phase separation is observed at all densities  for sufficiently high activity. The extent of phase separation is more at high density. As the temperature of hot dumbbells is 
increased, the phase separated cold dumbbells form a cluster with crystalline order while the hot dumbbells are in gaseous phase. We also determine the critical activity  from the distribution of parameter \(\psi\) defined in equation 9 and find that the critical activity of dumbbells is higher than that of Lennard-Jones monomer. The critical activity also increases with decrease in density. The entropy of hot dumbbells in the non-equilibrium system increases with the activity due to increase in effective volume on phase separation. The  density of cold dumbbells is greater than average density of the system while the density of hot dumbbells is less than the average density of the system . The crystalline nature of cluster of cold dumbbells is studied by bond-orientation order parameter and is found to be predominantly fcc and hcp. We also find that critical activity decreases as we increase the fraction of hot dumbbells.
The representation of 'activity' by temperature helps in understanding the physics of non-equilibrium systems.

\section*{Conflicts of interest}
There are no conflicts to declare.



\balance


\bibliography{rsc}
\bibliographystyle{rsc} 
\newpage

\section*{Supplementary Information}
\subsection*{1.Comparison with the equilibrium system}
Even though, the hot and cold subsystem were maintained at fixed temperatures at each simulation run, we see that the temperature varies as we move along interface even within the hot subsystem of the phase separated system(See Fig:12 in the main article). So we have plotted the velocity distribution of the whole hot subsystem and compared it with velocity distribution of hot dumbbells lying in the middle of phase separated hot subsystem.  In Fig:S1, the red plot shows velocity distribution of the whole hot subsystem and the blue plot shows velocity distribution of the hot dumbbells within the  hot sub system(away from interface) in range x/L=0.5 to x/L=0.7(refer Fig:11 and 12 in main article) when  T\textsuperscript{h*}=80 and density \(\rho\)*=0.8. The velocity distribution of the whole hot sub-system(red) is clearly not Maxwellian. But the velocity distribution of hot dumbbells away from interface (blue) resembles a Maxwellian distribution.

\begin{figure}[!h]
\centering
\includegraphics[width=1\columnwidth]{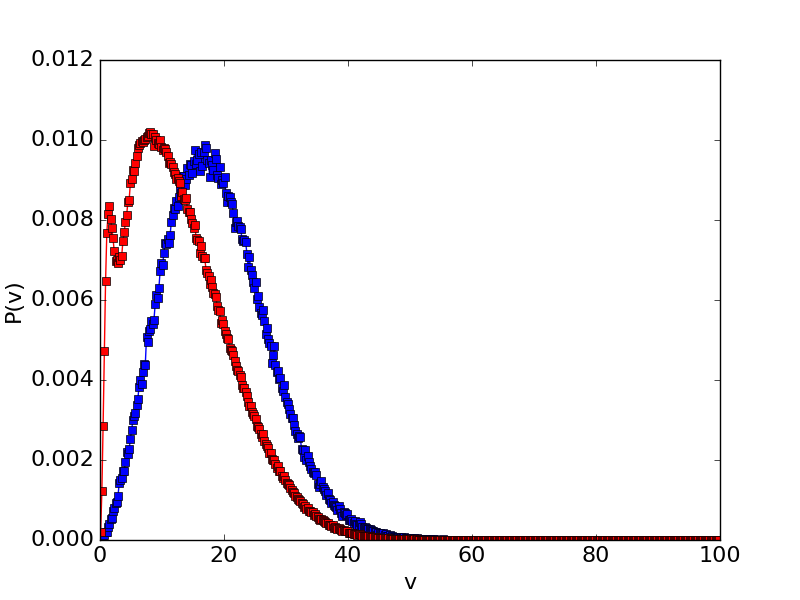}
\caption*{S1: The red plot shows velocity distribution of the whole hot subsystem and the blue plot shows velocity distribution of the hot dumbbells within the  hot sub system in range x/L=0.5 to x/L=0.7 when  T\textsuperscript{h*}=80.} 
\end{figure}
We have the density, temperature and pressure of the sub-volumes along the x-axis(refer Fig:11, 12, 13 in main article) at different T\textsuperscript{h*}. So we choose few sub-volumes, some of which are near interface and the rest away from interface and compare the pressure of the sub-volumes with the pressure obtained in a equilibrium system which has the same density and temperature  as that of the sub-volume.

In Supplementary Table 1, x/L refers to the position of the sub-volume, whose density(\(\rho\)*), temperature(T*) and pressure(P*(neq)) is given. x/L =0.3 indicates the sub-volume is located near interface. x/L=0.6 indicates sub-volume is present away from interface within the phase separated hot subsystem. P*(eq) refers to the pressure of an equilibrium system with the same density and temperature as that of corresponding sub-volume. \(\Delta\)P refers to the difference between P*(neq) and P*(eq). We can see from the table that for sub-volume near interface(x/L=0.3) the difference \(\Delta\)P is larger compared to sub-volumes away from interface(x/L=0.6). So as we move away from the interface, the sub-volumes behave more closely to their equilibrium counter-parts.
\begin{table}[!h]
    \centering
    \begin{tabular}{|c|c|c|c|c|c|c| }
      \hline
 x/L & T\textsuperscript{h*}& \(\rho\)* & T* & P*(neq)&P*(eq)&\(\Delta\)P\\  \hline
 0.3 & 80 & 0.80 & 32.46 & 57.85 &59.45&1.6\\
  0.3 & 60&0.78 & 26.22 & 45.93 &47.64&1.71\\ 
 0.3 & 40&0.78 & 17.69 & 33.90&35.00&1.1\\ 
 0.3 & 20&0.87 & 6.34 & 19.01 &20.71&1.7\\ 
  0.6 & 80&0.43 & 139.09 & 58.64 &58.21&0.43\\ 
 0.6 & 60&0.44 & 107.17 & 47.47 &47.88&0.41\\
  0.6 & 40&0.44 & 73.75 & 34.27 &33.82&0.45\\
   0.6 & 20&0.47 & 35.36 & 20.1 &19.77&0.33\\

 \hline    
    \end{tabular}
    \caption*{Supplementary Table 1: Here pressure of non-equilibrium system P*(neq) is compared with pressure of corresponding equilibrium system P*(eq). We can see from the table that for sub-volume near interface(x/L=0.3) the difference \(\Delta\)P is larger compared to sub-volumes away from interface(x/L=0.6).}
    \label{tab:my_label}
\end{table}
\subsection*{2. Effect of thermostat parameters on simulation.}
In section 2 in the main article, it is mentioned that Nos\'e-Hoover thermostat is used to maintain the temperature of dumbbells. The damping parameter T-damp is 50\(\delta\)t, where \(\delta\)t is the time step used in simulation. T-damp determines how quickly the temperature of the system relaxes to temperature value specified by the thermostat. Larger the value of T-damp, slower the system reaches the temperature specified by the thermostat. In the non-equilibrium system, for cold subsystem the value of T-damp determines the rate at which it removes the energy transferred from the hot subsystem. To study the effect of T-damp we perform simulation of the non-equilibrium system at \(\rho\)* = 0.8 varying the damping parameter T-damp=50\(\delta\)t, 100\(\delta\)t, 200\(\delta\)t. In Fig:S2, plot of effective temperature  of cold dumbbells \(T^{c*}_{eff}\) versus temperature of hot dumbbells  \(T^{h*}\) and plot of activity \(\chi\) versus temperature of hot dumbbells \(T^{h*}\) at different T-damp is given. We can see that as the value of T-damp is increased, the effective temperature of cold dumbbells  \(T^{c*}_{eff}\) also increases. This is because large value of T-damp implies a slow rate at which energy is removed from cold system as a result of which the cold-dumbbells reach steady state at high effective temperature. The increase in \(T^{c*}_{eff}\) with T-damp results in decrease in activity \(\chi\) with increase in T-damp. So the value of T-damp determines the activity of the non-equilibrium system for given values of T\textsuperscript{c*} and T\textsuperscript{h*}.
\begin{figure*}[h]
\centering
\includegraphics[width=1\textwidth]{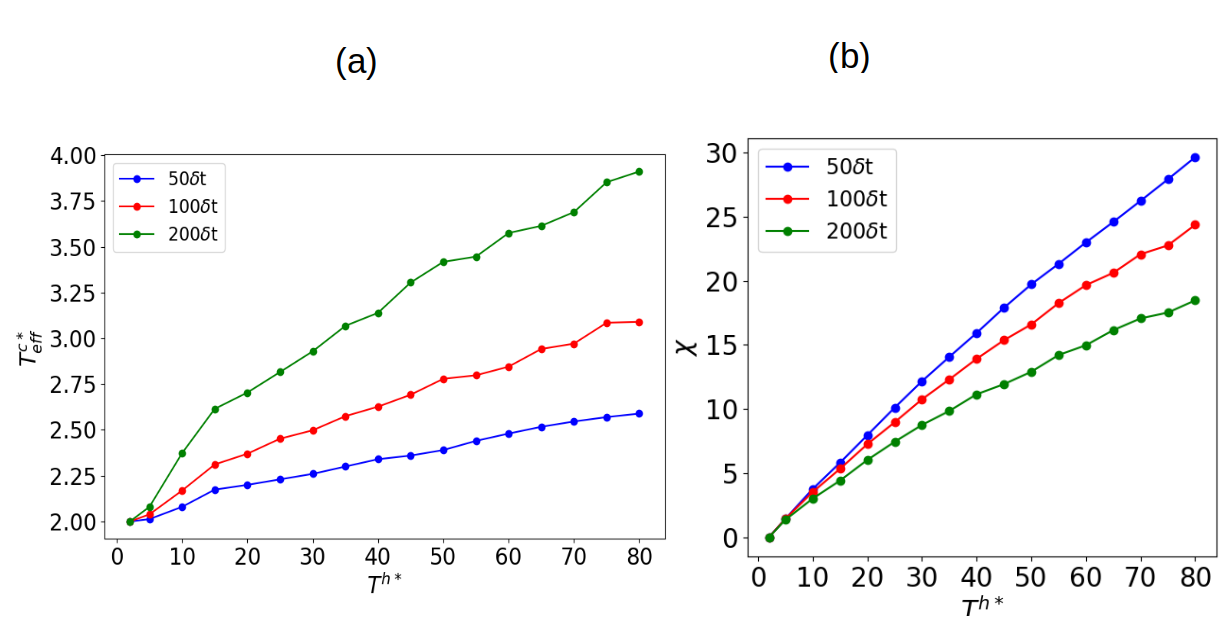} 
\caption*{S2: a)Plot of effective temperature of cold dumbbells \(T^{c*}_{eff}\) versus \(T^{h*}\) at different T-damp. b) Plot of activity \(\chi\) versus \(T^{h*}\) at different T-damp. We can see that \(T^{c*}_{eff}\) increases with increase in T-damp. Conversely,activity \(\chi\) decreases with increase in T-damp.}
\end{figure*}

\subsection*{3.Additional Supplementary figures}

\begin{figure*}[!h]
\centering
\includegraphics[width=1\textwidth]{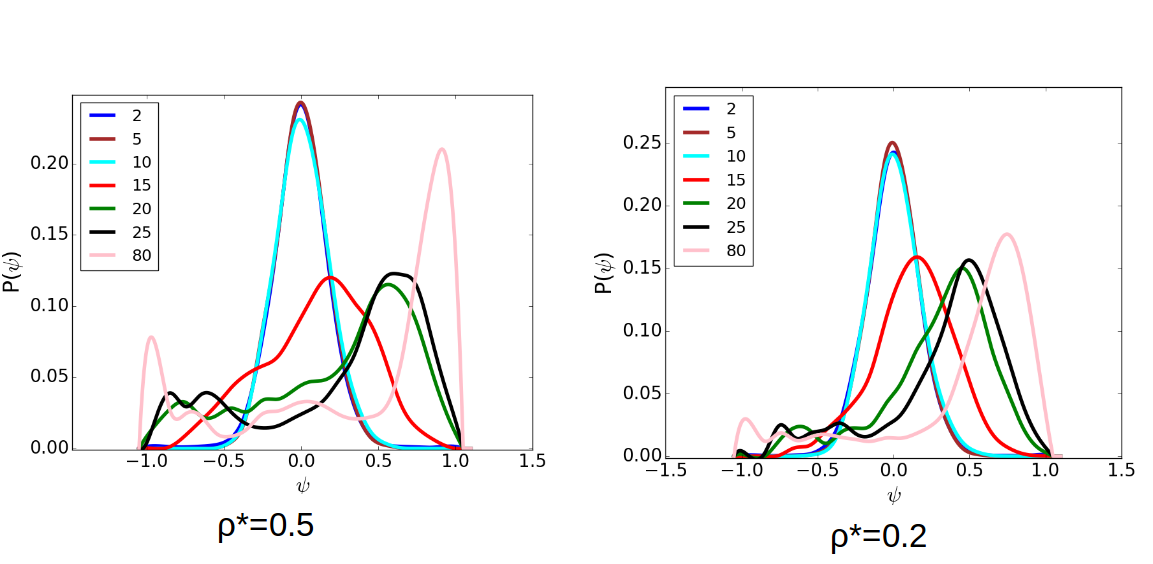} 
\caption*{S3: Probability distribution P(\(\psi\)) of \(\psi\) at density \(\rho*\)=0.5 and 0.2 for different values of temperature of hot dumbbells T\textsuperscript{h*}. The bimodality in distribution P(\(\psi\)) of  \(\psi\) occurs at T\textsuperscript{h*}=20  for both density \(\rho*\)=0.5 and 0.2 respectively. These temperature of hot dumbbells T\textsuperscript{h*} correspond to the critical activity \(\chi\) of 8.41 and 8.74 respectively.}
\end{figure*}

\begin{figure*}[!h]
\includegraphics[width=0.7\textwidth]{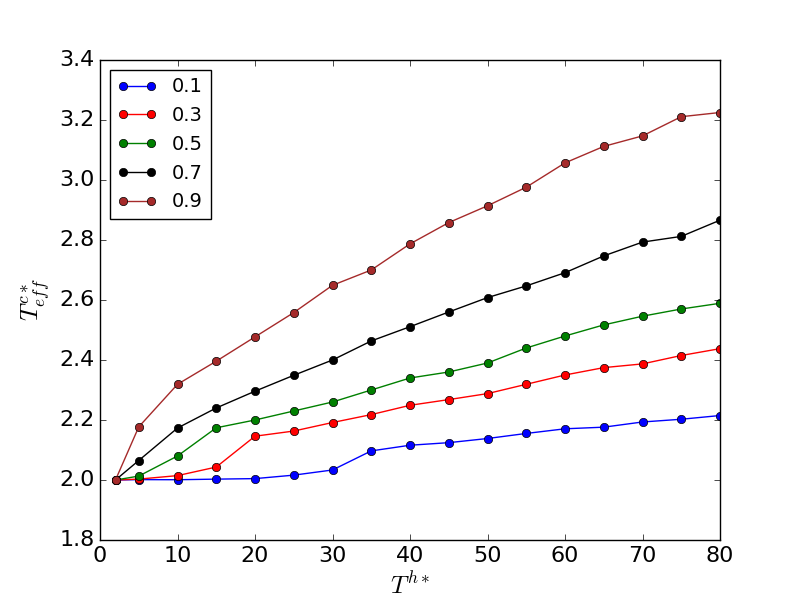} 
\caption*{S4: Plot of effective temperature of cold dumbbells \(T_{eff}^{c*}\) versus temperature of hot dumbbells T\textsuperscript{h*} at density \(\rho*\)=0.8 for different fractions of hot dumbbells f\textsubscript{h}. We can see that as the fraction of hot dumbbells f\textsubscript{h} increases the effective temperature of cold dumbbells  \(T_{eff}^{c*}\) also increases.  }
\end{figure*}
\begin{figure*}[!h]
\includegraphics[width=0.7\textwidth]{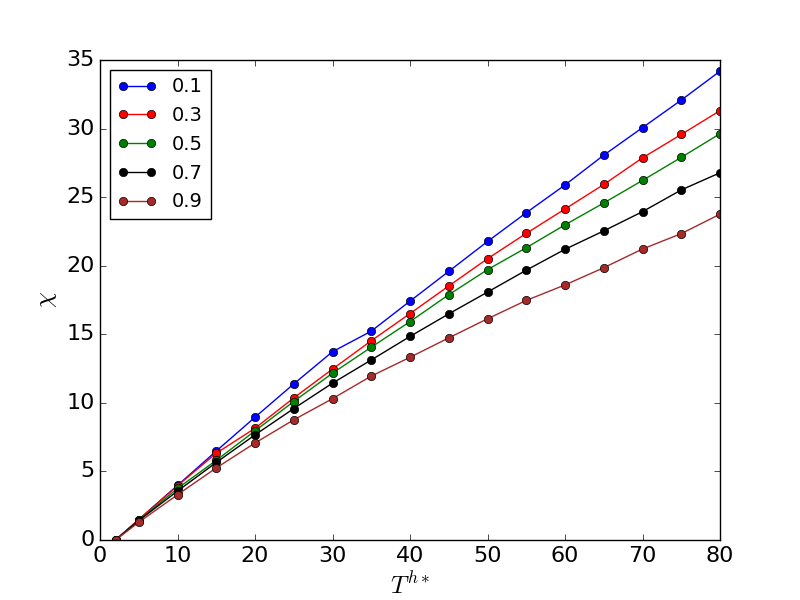} 
\caption*{S5: Plot of activity \(\chi\) versus temperature of hot dumbbells T\textsuperscript{h*} at density \(\rho*\)=0.8 for different fractions of hot dumbbells f\textsubscript{h}. We can see that as the fraction of hot dumbbells f\textsubscript{h} increases the activity \(\chi\) decreases.  }
\end{figure*}
\begin{figure*}[!h]
\includegraphics[width=0.7\textwidth]{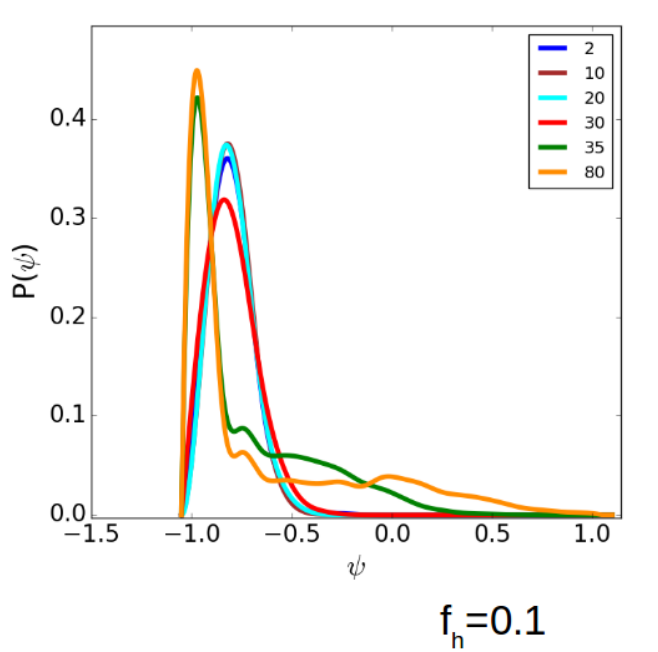} 
\caption*{S6: Probability distribution P(\(\psi\)) of \(\psi\) at density \(\rho*\)=0.8 for  fraction of hot dumbbells f\textsubscript{h}=0.1  at different T\textsuperscript{h*}. Due to unequal number of hot and cold dumbbells the distribution tends to be towards the majority species.  The bimodality in distribution P(\(\psi\)) of  \(\psi\) occurs at T\textsuperscript{h*}=35   for f\textsubscript{h}=0.1. The temperature of hot dumbbells T\textsuperscript{h*}=35 corresponds to activity \(\chi\) of 15.28.  }
\end{figure*}

\begin{figure*}[!h]
\includegraphics[width=0.7\textwidth]{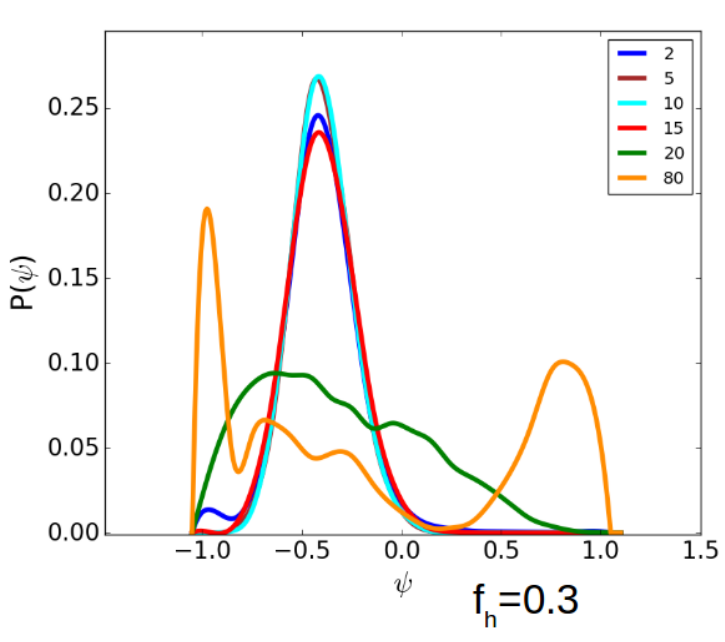} 
\caption*{S7: Probability distribution P(\(\psi\)) of \(\psi\) at density \(\rho*\)=0.8 for  fraction of hot dumbbells f\textsubscript{h}=0.3  at different T\textsuperscript{h*}. Due to unequal number of hot and cold dumbbells the distribution tends to be towards the majority species.  The bimodality in distribution P(\(\psi\)) of  \(\psi\) occurs at T\textsuperscript{h*}=20   for f\textsubscript{h}=0.3. The temperature of hot dumbbells T\textsuperscript{h*}=20 corresponds to activity \(\chi\) of  8.14.  }
\end{figure*}

\begin{figure*}[!h]
\includegraphics[width=0.7\textwidth]{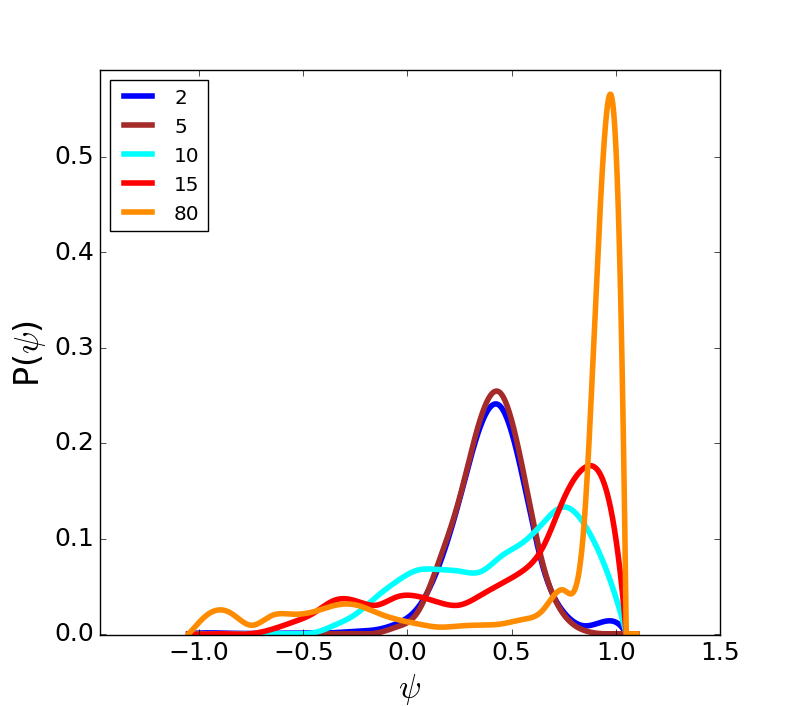} 
\caption*{S8: Probability distribution P(\(\psi\)) of \(\psi\) at density \(\rho*\)=0.8 for  fraction of hot dumbbells f\textsubscript{h}=0.7 at different T\textsuperscript{h*}. Due to unequal number of hot and cold dumbbells the distribution tends to be towards the majority species.  The bimodality in distribution P(\(\psi\)) of  \(\psi\) occurs at T\textsuperscript{h*}=10 which corresponds to activity \(\chi\) of 3.56.  }
\end{figure*}

\begin{figure*}[!h]
\includegraphics[width=0.7\textwidth]{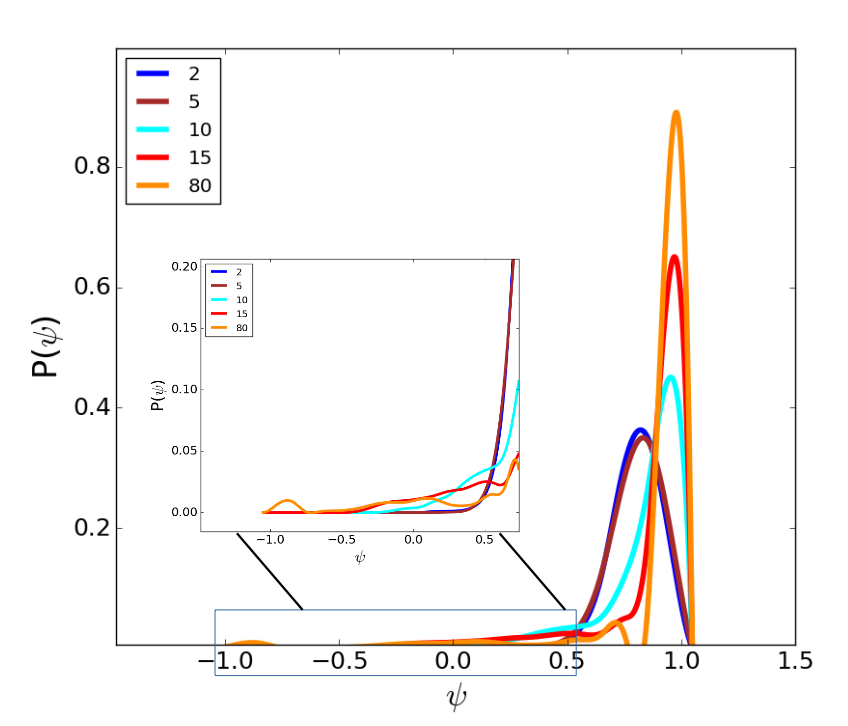} 
\caption*{S9: Probability distribution P(\(\psi\)) of \(\psi\) at density \(\rho*\)=0.8 for  fraction of hot dumbbells f\textsubscript{h}=0.9 at different T\textsuperscript{h*}. Here since the hot dumbbells exceedingly dominate cold-dumbbells. So we have magnified the region -1\(<\psi<\)0.5.  The bimodality in distribution P(\(\psi\)) of  \(\psi\) occurs at T\textsuperscript{h*}=10 which corresponds to activity \(\chi\) of 3.30.  }
\end{figure*}

\end{document}